\documentclass[twocolumn,amsfonts,groupedaddress,nofootinbib]{revtex4-1}

\usepackage{amsmath}
\usepackage{color}
\usepackage{graphicx}
\usepackage{subfigure}
\usepackage{amssymb}
\usepackage{hyperref}

\pdfoutput=1

\def\bea#1\eea{\begin{align}#1\end{align}}
\def\be#1\ee{\begin{equation}#1\end{equation}}

\DeclareFontFamily{OT1}{pzc}{}
\DeclareFontShape{OT1}{pzc}{m}{it}{<-> s * [1.1] pzcmi7t}{}
\DeclareMathAlphabet{\mathpzc}{OT1}{pzc}{m}{it}

\newcommand{\D}{\mathpzc{D}}
\newcommand{\T}{\mathcal{T}}
\newcommand{\Hcal}{\mathpzc{H}}
\newcommand{\Lcal}{\mathpzc{L}}
\newcommand{\Rcal}{\mathpzc{R}}
\newcommand{\Ocal}{\mathcal{O}}

\newcommand{\gcal}{\mathpzc{g}}
\newcommand{\rhob}{\bar{\rho}}
\newcommand{\phib}{\bar{\phi}}
\newcommand{\gb}{\bar{g}}
\newcommand{\Dbar}{\overline{D}}
\newcommand{\Rbar}{\overline{R}}
\newcommand{\Lbar}{\overline{L}}

\begin{document}


\title{Bounds on the local energy density of holographic CFTs from bulk geometry}

\author{Sebastian Fischetti}
\email[]{s.fischetti@imperial.ac.uk}
\author{Andrew Hickling}
\email[]{a.hickling12@imperial.ac.uk}
\author{Toby Wiseman}
\email{t.wiseman@imperial.ac.uk}
\affiliation{Theoretical Physics Group, Blackett Laboratory, Imperial College, London SW7 2AZ, UK}

\date{April 2016}

\begin{abstract}

The stress tensor is a basic local operator in any field theory; in the context of AdS/CFT, it is the operator which is dual to the bulk geometry itself.  Here we exploit this feature by using the bulk geometry to place constraints on the local energy density in static states of holographic~$(2+1)$-dimensional CFTs living on a closed (but otherwise generally curved) spatial geometry.  We allow for the presence of a marginal scalar deformation, dual to a massless scalar field in the bulk.  For certain vacuum states in which the bulk geometry is well-behaved at zero temperature, we find that the bulk equations of motion imply that the local energy density integrated over specific boundary domains is negative.  In the absence of scalar deformations, we use the inverse mean curvature flow to show that if the CFT spatial geometry has spherical topology but non-constant curvature, the local energy density must be positive somewhere.  This result extends to other topologies, but only for certain types of vacuum; in particular, for a generic toroidal boundary, the vacuum's bulk dual must be the zero-temperature limit of a toroidal black hole.
\end{abstract}

\maketitle

%
\section{Introduction}
\label{sec:intro}
%

AdS/CFT \cite{Maldacena:1997re,Gubser:1998bc,Witten:1998qj} provides a fascinating way to understand the behaviour of specific strongly coupled conformal field theories.  It has the potential to be a powerful tool to better understand theories of physical interest, such as QCD~\cite{CasalderreySolana:2011us} or condensed matter at strongly coupled critical points~\cite{Hartnoll:2009sz}.  However, since it is unclear whether or not holographic gauge theories describe our universe, an important goal is to understand what lessons may be gleaned from them that apply to non-holographic systems more directly relevant to our universe.  To this end, it is interesting to focus on behaviour that is universal amongst holographic models and also universal to classes of states described by these models. Such lessons have the potential to generalise to other strongly coupled non-holographic theories, whereas detailed model or state dependent phenomena are presumably less likely to do so.  It is this search for universal behaviours in theories with a gravity dual that is the motivation for this work.

One of the most basic operators in any CFT (indeed, in any QFT) is the stress tensor; in AdS/CFT, this object is intimately linked to the existence of a gravitational dual.  All holographic models described by usual two-derivative semiclassical gravitational bulk theories have a ``universal sector'' described by pure Einstein gravity in an asymptotically locally AdS spacetime.  This sector is dual to the dynamics of the CFT stress tensor, and is the reason the stress tensor plays a special role.  In this work we consider a~$(2+1)$-dimensional holographic CFT and this universal sector.  We will restrict to the class of states which are static and in which the CFT has been placed on a spacetime which is the ultrastatic product of time with a closed but otherwise general spatial geometry (i.e.~a Riemann surface).  In particular, in this static context this singles out the local energy density from the stress tensor as a basic observable of interest; this the quantity on which we will focus in this paper.  We should note that there has been considerable interest in constraints on the behaviour of the local energy density of QFTs in the study of ``quantum energy inequalities'' (QEIs) \cite{For09,Few12} which typically bound integrals of the local energy density from above.  Such bounds are hard to derive even for the simplest case of free fields\footnote{Though some progress has been made for e.g.~the non-minimally coupled scalar field~\cite{FewOst07}.}, and obtaining analogous bounds in general strongly coupled theories on curved spacetimes directly seems an intractable challenge.  This is again motivation for why constraints on the stress tensor in classes of strongly coupled theory with gravity duals are potentially so interesting~\cite{Marolf:2013ioa}.

To keep our results slightly more general, in the first portion of this paper we will in fact also allow deformations of the CFT by a static but otherwise general marginal CFT scalar operator, which is dual to a massless scalar in the bulk\footnote{The gravity and massless scalar sector arises as a consistent top-down reduction in the $(3+1)$-dimensional holographic CFT setting, in which case the scalar may be taken to be the Type IIB dilaton and is associated to the existence of the exactly marginal gauge coupling.  In the $(2+1)$-dimensional setting it is less clear that such massless fields arise in consistent truncations; for example, in the case of reductions of 11-dimensional supergravity on~AdS$_4$ one does not necessarily obtain these~\cite{Gauntlett:2009zw}. 
}.
Thus the space of deformations considered is the space of smooth functions over Riemann surfaces. 

We derive two classes of results.  Firstly, for static thermal states dual to Einstein-scalar gravity we consider a certain one-parameter family of regions~$U_\zeta$ of the CFT geometry which are defined in terms of the curvature and scalar source.  We show that the energy of these regions is bounded above by objects constructed from the geometry and scalar field on any bulk horizons.  This upper bound vanishes in the limit of zero temperature, so that for static vacuum (zero-temperature) CFT states, the energy of the regions~$U_\zeta$ is generically negative (and may vanish only when the CFT spatial geometry and scalar source are homogeneous).  This result generalises that of~\cite{HicWis15b}: for pure gravity, the total energy -- the Casimir energy -- of the CFT is non-positive (see also the earlier result~\cite{Galloway:2015ora} for fixed homogeneous boundary space).  Below we will make precise how the~$T \to 0$ limit should behave; here we note only that the zero temperature bulk solution is not required to be smooth, but may have singularities of a ``good'' variety~\cite{Gubser:2000nd}. 

This first class of results raises the question of whether the energy density must be everywhere negative for such vacua.  Our second class of results provides an answer to this question.  Restricting now to the universal sector (i.e.~pure Einstein gravity in the bulk with arbitrary boundary spatial geometry), we show certain integrals of the energy density are bounded \textit{below} by geometric data at bulk horizons.  Restricting to zero temperature static vacuum states, these bounds imply that for a boundary with spherical topology the energy cannot be negative everywhere.  For a boundary with genus~$\mathfrak{g} \geq 1$, we obtain the same result provided the vacuum is one such that raising it to any small but finite temperature results in the bulk ending on a horizon with the same topology as the boundary (and additionally which for~$\mathfrak{g} > 1$ has vanishing entropy in the zero-temperature limit).

We emphasise that both sets of results derive from relatively simple geometric considerations. For the first results we use simple Riemannian methods.  Our second results derive from a simple generalisation of the work of Lee and Neves~\cite{LeeNev15} on Inverse Mean Curvature (IMC) flow in hyperbolic spaces. It is likely that more sophisticated geometric techniques could provide stronger constraints on physical quantities that derive from the existence of a dual gravity description.

This paper is structured as follows. In Section~\ref{sec:results} we provide a detailed summary of our results.  In Section~\ref{sec:bulk} we review some basic properties of the bulk which will then be used in Section~\ref{sec:negativeE} to derive the first class of results mentioned above.  In Section~\ref{sec:IMCF} we introduce the IMC flow and use it to derive constraints on the negativity of the local energy density.  We conclude in Section~\ref{sec:disc} with a discussion of interesting potential generalisations of our results and future directions.

\noindent \textbf{Notation:} For the benefit of the reader, here we summarise the notation used in this paper.  Tensorial objects in the bulk spacetime will be dressed with a left superscript~$(4)$ (e.g.~$^{(4)} \! g_{ab}$,~$^{(4)} \! R_{ab}$, etc.), while tensorial objects living in the optical geometry (defined under~\eqref{eq:optical} below) of a static bulk time slice will be undecorated (e.g.~$g_{ab}$,~$R_{ab}$, etc.).  Boundary quantities will be dressed with an overline; tensorial objects living on a static boundary spatial slice will be given no other decoration (e.g.~$\gb_{ab}$,~$\Rbar_{ab}$), while tensorial objects living in the full Lorentzian boundary geometry will be given a left superscript~$(3)$ (e.g.~$^{(3)} \! \gb_{ab}$,~$^{(3)} \! \Rbar_{ab}$, etc.).  Objects living on the bifurcation surface of a bulk horizon will be typeset in calligraphic font (e.g.~$\gcal_{ab}$,~$\Rcal_{ab}$, etc.).  We will use a capital~$D$ (dressed appropriately) for covariant derivatives on these geometries.  Less-frequently used notation (specifically in Section~\ref{sec:IMCF}) will be defined as it is introduced.  

\noindent \textbf{Index Conventions:} Lower-case letters from the beginning of the Latin alphabet ($a,b,\ldots$) will be used as abstract indices; bulk spacetime coordinates will be indexed by upper-case letters from the beginning of the Latin alphabet~($A,B,\ldots$); boundary spacetime coordinates will be indexed by lower-case letters from the middle of the Greek alphabet ($\mu,\nu,\ldots$); bulk spatial coordinates will be indexed by lower-case letters from the middle of the Latin alphabet ($i,j,\ldots$); and two-dimensional spatial coordinates (used for either spatial slices of the boundary or a horizon) will be indexed by lower-case letters from the beginning of the Greek alphabet ($\alpha,\beta,\ldots$).

%
\section{Summary of Results}
\label{sec:results}
%

In this section we summarise our results, beginning by setting the framework in which we work.  We are interested in a static state of a CFT living on a (globally static) boundary metric~$^{(3)} \! \gb_{ab}$ whose spatial geometry~$(\partial M, \gb_{ab})$ is closed but generally curved\footnote{We emphasise that here we use the notation~$\partial M$ to denote only a static time slice of the CFT spacetime, not the entire spacetime itself.}.  Unless otherwise specified, we will also turn on a (static) boundary source~$\phib$ for the bulk scalar.  Thus we assume there exists a static infilling bulk solution whose time-translation Killing vector field becomes that of the boundary\footnote{More precisely, the conformal Killing vector of the boundary conformal class.} when suitably restricted.  As usual, we introduce a time coordinate~$t$ adapted to the static isometry such that~$(\partial_t)^a$ is the time-translation Killing vector field of both the bulk and boundary.

We now consider two cases: either the CFT is in a thermal state with nonzero temperature~$T$, or it is in a vacuum state with zero temperature.  Note that when we refer to a vacuum state from this point on it will be implicit that we are referring to zero temperature. 

If the CFT is in a thermal state at (nonzero) temperature~$T$, we assume that in addition to the conformal boundary, the bulk geometry may also end on some number of regular horizons, each with the same surface gravity~$\kappa = 2 \pi T$ (with respect to~$(\partial_t)^a$).  If there are no bulk horizons we expect the CFT is in a confined phase, and with bulk horizons in a deconfined phase~\cite{Witten:1998zw}.

If instead the CFT is in a (zero-temperature) vacuum state, we follow \cite{HicWis15b} and assume the bulk metric is the~$T \to 0$ limit of some finite temperature solution.  In fact, we will restrict to three varieties of vacuum geometries, classified according to the presence of horizons and the behaviour of their area~$A_\Hcal$ in the zero-temperature limit:
\begin{enumerate}
\item {\bf Confined vacuum}: the bulk has no ends other than the asymptotic AdS boundary.
\item {\bf Deconfined vacuum}: besides the asymptotic AdS boundary, the bulk ends on a null surface that is the $T \to 0$,~$A_\Hcal \to 0$ limit of a black hole horizon \emph{with the same topology} as the boundary.
\item {\bf Degenerate vacuum}: besides the asymptotic AdS boundary, the bulk ends on an extremal horizon which is the $T \to 0$ but~$A_\Hcal \nrightarrow 0$ limit of a black hole horizon \emph{with the same topology} as the boundary.
\end{enumerate}
We note here that the deconfined and degenerate vacua may contain more than one horizon or null surface; the above categories only require at least one of these to have the same topology as the boundary.

While there may exist vacua which do not fall into any of these categories, all the vacua of which we are aware do.  For instance, global AdS and the AdS soliton~\cite{HorMye98} are examples of confined vacua with spherical and toroidal boundary, respectively.  Compact quotients of the extremal hyperbolic AdS-Schwarzschild black hole~\cite{Emp99} provide examples of degenerate vacua in which the genus~$\mathfrak{g}$ of the boundary (and extremal horizon) is greater than one.  Finally, we may obtain a deconfined vacuum by quotienting the Poincar\'e patch of pure AdS so that the boundary geometry becomes a flat torus.  As emphasised in \cite{HicWis15b}, this is an important physical example: while the Poincar\'e patch ends on a smooth extremal horizon, when it is compactified to a spatial torus the IR geometry becomes a null singularity~\cite{Kunduri:2013ana}.  Nevertheless, it is a ``good'' singularity in the sense that at any nonzero temperature the bulk solution becomes planar AdS-Schwarzschild which has a perfectly regular horizon when compactified to a torus.

We remind the reader of the basis for this terminology.  In a confined vacuum, we expect that since the only end is the conformal boundary then the fluctuation spectrum will be gapped rather than continuous, with the gap representing the confinement scale which in turn is set by a geometric scale in the boundary data.  In addition the geometry will not change in the supergravity approximation going to small finite temperature, and hence the entropy associated to thermal excitation about such a vacuum will be~$O(1)$, which is characteristic of confinement in a large~$N$ gauge theory.  On the other hand, in a deconfined vacuum we expect the entropy to grow as~$O(N^2)$ when finite temperature is introduced (corresponding to the presence of a horizon), and the fluctuation spectrum is expected to be continuous due to the presence of the extended null surface of infinite redshift in the IR of the bulk vacuum geometry.  Finally, the non-vanishing area of an extremal horizon implies that the entropy of the dual CFT state is~$O(N^2)$, and thus such a vacuum state is highly degenerate.

Next, since the static Killing vector field~$(\partial_t)^a$ of $^{(3)} \! \gb_{ab}$ is globally timelike, we may work in an ultrastatic conformal frame where~$\|\partial_t\|^2 = -1$.  In this frame, the boundary metric takes the form
\be
\label{eq:boundary}
^{(3)} \! \gb_{\mu\nu} dx^\mu dx^\nu = -dt^2 + \gb_{\alpha\beta}(x) dx^\alpha dx^\beta,
\ee
while the local energy density~$\rhob = \langle T_{tt} \rangle $.  Note that while we will cast our results in terms of~$\rhob$, we can re-express them in terms of the local energy density~$\rho$ in any other frame using~$\rhob = (-\|\partial_t\|^2)^{3/2} \rho$.

Now, consider the tensor
\begin{align}
\label{eq:Lbar}
\Lbar_{ab} \equiv \Rbar_{ab} -  \left(\Dbar_a \phib\right) \left(\Dbar_b \phib\right)
\end{align}
defined on a spatial slice~$(\partial M, \gb_{ab})$ of the ultrastatic geometry (and recall that~$\Rbar_{ab}$ is the Ricci tensor of $\gb_{ab}$).  Now, there must exist points $p^\star$ on the boundary~$\partial M$ where the trace $\Lbar \equiv \gb^{ab} \Lbar_{ab}$ is (globally) minimised.  We define the regions containing these points and bounded by a level set~$\Lbar = \zeta$ for some constant~$\zeta$ as
\be
\label{eq:Szeta}
U_\zeta \equiv \{ p \in \partial M | \Lbar(p) < \zeta \}.
\ee
Denoting the energy of such a region as
\be
\label{eq:Ezeta}
E_\zeta \equiv \int_{U_\zeta} \sqrt{\gb} \, \rhob,
\ee
then our first result is that for any value of $\zeta > \min(\Lbar)$, 
\begin{align}
\label{eq:result1}
\mathrm{Result\;1:} \quad E_\zeta \le 0
\end{align}
in any of the three vacua we consider\footnote{In fact, this result holds in the deconfined and degenerate vacua even if the topology constraint is removed.}.  Moreover, equality is only possible if~$\Rbar$ and~$\phib$ are constant (which in turn implies~$\Lbar$ is constant as well); in this case,~$U_\zeta$ is either empty (if~$\zeta \leq \Lbar$) or the entire boundary (if~$\zeta > \Lbar$).  Thus it is only possible to obtain an equality in~\eqref{eq:result1} on the total energy~$E$ of the CFT.

A corollary of~\eqref{eq:result1} is that if~$\Lbar$ attains its global minimum at only a single point~$p^\star$, then the local energy density at this point must be non-positive:
\begin{align}
\label{eq:result1cor}
\rhob(p^\star) \leq 0;
\end{align}
in addition, we show that this inequality becomes strict in the confined vacuum.
 
This result generalises that in \cite{HicWis15b} where the total energy of the vacuum was found to be non-positive when there was pure gravity in the bulk. Interestingly, in that case the bound was related to geometric inequalities involving the Gauss-Bonnet theorem and the renormalised bulk volume~\cite{Anderson}.  In contrast, here we have included a massless scalar, allowed deformations to its dual, and have a local version of the bound rather than just a global statement about the total energy.

Given that $E_\zeta \le 0$ for any $\zeta$, a natural question is whether in fact $\rhob$ must be non-positive everywhere.  We will provide an answer to this question when boundary scalar deformations are turned off, which will imply that~$\phi$ in the bulk is constant and hence can be ignored, effectively leaving pure gravity.  Then using a simple adaptation of the rigorous geometric result in~\cite{LeeNev15}, we come to the interesting conclusion that in many cases~$\rhob$ cannot be everywhere negative: for a sphere topology boundary deformed away from being round, in any of the three vacua we must have~$\rhob > 0$ somewhere on the boundary.  For a higher genus~$\mathfrak{g} > 0$ topology boundary with non-constant scalar curvature, this result only generalises to some of the vacuum classes:
\begin{align}
\label{eq:result2}
\begin{array}{c}
\mbox{Result 2:} \\
\left. \begin{array}{c}
\mathfrak{g} = 0 \\ \mbox{(non round)}, \\ \mbox{any vacuum} \\
\\
\mathfrak{g} = 1 \\ \mbox{(non flat)}, \\ \mbox{deconfined/degenerate} \\
\\
\mathfrak{g} > 1 \\ \mbox{(non const. curv.)}, \\ \mbox{deconfined} 
\end{array} \right\} 
 \implies \begin{array}{c} 
\exists \, p \in \partial M \; \mathrm{so} \\
\rhob(p) > 0
\end{array}
\end{array}
\end{align}
We note that for pure Einstein gravity in the bulk, we are unaware of any examples of deconfined vacua with~$\mathfrak{g} > 1$ or degenerate vacua with~$\mathfrak{g} \leq 1$.  Indeed, Result~2 may be interpreted as a constraint on which classes of vacuum may exist given the positivity properties of~$\rhob$.  Specifically, the vacua listed in Result~2 cannot exist if~$\rhob$ is everywhere non-positive.  For the special case of spherical topology~$\mathfrak{g} = 0$, this result implies that any CFT vacuum state must have~$\rhob > 0$ somewhere.

Finally, we note that for genus~$\mathfrak{g} = 0$ or~$1$, Result~2 admits an extension to nonzero temperature, subject to the constraint that for~$\mathfrak{g} =1$ there exists a toroidal horizon in the bulk. For higher genus~$\mathfrak{g} > 1$, we find an analogous result provided the temperature is sufficiently large,~$T^2 \ge -\min(\Rbar)/(8 \pi^2)$, and there exists a horizon of genus~$\mathfrak{g}$ in the bulk.

%
\section{The Dual Bulk Geometry}
\label{sec:bulk}
%

In this Section, we outline some properties of the bulk that will be useful in the derivation of our results.  Let the $(3+1)$-dimensional bulk metric be~$^{(4)} \! g_{ab}$ and the massless bulk scalar be~$\phi$.  Then the full bulk equations of motion are
\begin{subequations}
\bea
^{(4)} \! D^2 \phi & = 0, \\
^{(4)} \! R_{ab} -  \left(\, ^{(4)} \! D_a \phi \right)\left(\, ^{(4)} \! D_b \phi\right) &= - \frac{3}{\ell^2} \, ^{(4)} \! g_{ab} ,
\eea
\end{subequations}
where the AdS scale~$\ell$ is related to the CFT ``effective central charge'' as~$c = \ell^{2}/(16 \pi G_{(4)})$, with $G_{(4)}$ the bulk Newton constant.

\subsubsection{Near-Horizon}

Now, consider some (connected) bulk Killing horizon component~$\Hcal$.  Locally near~$\Hcal$ we can introduce a normal coordinate~$r$ to write the metric as
\be
ds^2 = -\kappa^2 r^2 Q(r,x) dt^2 + dr^2 + \, ^\Hcal \! g_{\alpha\beta}(r,x) dx^\alpha dx^\beta,
\ee
where the horizon is at~$r = 0$ and~$\kappa$ is its (constant) surface gravity (this implies that we have normalised~$Q(0,x) = 1$).  Regularity then demands that~$Q$,~$\phi$, and the components~$^\Hcal \! g_{\alpha\beta}$ be smooth functions of $r^2$. The bulk equations locally determine these smooth functions in terms of the spatial geometry of the horizon (given by the 2-metric $\gcal_{\alpha\beta}(x) = \, ^\Hcal \! g_{\alpha\beta}(0,x)$) and the scalar profile $\varphi(x)$ on the horizon.  Expanding in $r^2$, the equations yield the expansions
\begin{subequations}
\label{eqs:horizon}
\bea
\phi(r,x) & = \varphi(x) - \frac{1}{4} \left(\D^2 \varphi \right) r^2  +  O(r^4), \label{eq:horizonscalar} \\
Q(r,x) &= 1 - \frac{1}{6} \Lcal r^2  +  O(r^4), \\
^\Hcal \! g_{\alpha\beta}(r,x) &= \gcal_{\alpha\beta} +  \frac{1}{2} \left( \Lcal_{\alpha\beta} + \frac{3}{\ell^2} \gcal_{\alpha\beta} \right) r^2  +  O(r^4),
\eea
\end{subequations}
where the terms in the expansion above are written covariantly in the 2-dimensional geometry~$\gcal_{\alpha\beta}$.  Note that as before, we have defined the tensor
\be
\Lcal_{\alpha\beta} \equiv \Rcal_{\alpha\beta} - \left(\D_\alpha \varphi \right) \left( \D_\beta \varphi \right),
\ee
where $\Rcal_{\alpha\beta}$ is the Ricci tensor of the horizon spatial metric $\gcal_{\alpha\beta}$.  In the semiclassical limit, the entropy of the horizon component is determined by its area~$A_\Hcal$ as~$S_\Hcal = 4 \pi c A_\Hcal/\ell^2$ and is summed with that of other horizon components to give the total CFT entropy at temperature~$T = \kappa/2\pi$.

\subsubsection{Near-Boundary}

Next, consider the asymptotically locally AdS boundary.  Recall that in an asymptotically locally AdS spacetime with metric~$^{(4)} \! g_{ab}$, the conformally rescaled metric~$\Omega^2 \, ^{(4)} \! g_{ab}$ induces a regular metric on the boundary for any~$\Omega$ satisfying~$\Omega \to 0$ and~$d\Omega \neq 0$ there~\cite{FisKel12}.  This function~$\Omega$ is referred to as a defining function, as it defines a conformal frame: that is, it determines a boundary metric which is a representative of the conformal class of the boundary.  Of course, the choice of~$\Omega$ is not unique; for any positive definite function~$f$, the choice~$\Omega' = f\Omega$ is a valid defining function as well.

Having chosen a conformal frame, we may then expand the bulk metric (and any bulk fields) near the boundary in Fefferman-Graham (FG) coordinates adapted to this frame; note in particular that the FG radial coordinate~$z$ obeys~$\lim_{z\to 0} \ell\Omega/z = 1$.  Working in the conformal frame in which the boundary metric takes the ultrastatic form~\eqref{eq:boundary}, we have\footnote{Recall that the time coordinate~$t$ is adapted to the static isometry, so no~$dt \, dx$ cross-terms may appear.}
\begin{multline}
^{(4)} \! g_{AB} dx^A dx^B = \frac{\ell^2}{z^2} \left( dz^2 + \, ^{(\mathrm{FG})} \! g_{tt}(z,x) \, dt^2 \right. \\ \left. + \, ^{(\mathrm{FG})} \! g_{\alpha\beta}(z,x) dx^\alpha dx^\beta \right),
\end{multline}
where~\cite{Skenderis:2002wp}
\begin{subequations}
\label{eqs:FG}
\be
^{(\mathrm{FG})} \! g_{tt}(z,x) = -1 - \frac{1}{4} \Lbar z^2 + \frac{\rhob}{3c} \, z^3 + O(z^4),
\ee
\begin{multline}
^{(\mathrm{FG})} \! g_{\alpha\beta}(z,x) = \gb_{\alpha\beta}(x) -  \left(\Lbar_{\alpha\beta} - \frac{1}{4} \gb_{\alpha\beta} \Lbar \right) z^2 \\ +  \frac{\left\langle T_{\alpha\beta}\right\rangle}{3c} \, z^3 + O(z^4),
\end{multline}
\be
\label{eq:FGscalar}
\phi(z,x) = \phib(x) +  \frac{1}{2} \left(\Dbar^2 \phib \right) z^2 + \frac{\left\langle \Ocal \right\rangle}{6c} z^3 + O(z^4).
\ee
\end{subequations}
Note that the terms at each order in~$z$ are written covariantly in the spatial boundary metric~$\gb_{\alpha\beta}$ of the ultrastatic frame, and recall that the tensor~$\Lbar_{\alpha\beta}$ was defined in~\eqref{eq:Lbar}.  Per the AdS/CFT dictionary, we have also written the expansion in terms of the one-point functions of scalar operators:~$\left\langle \Ocal \right\rangle$ is the one-point function of the CFT operator~$\Ocal$ dual to the bulk scalar, while~$\rhob$ and~$\left\langle T_{\alpha\beta} \right\rangle$ are the CFT energy density and spatial stress tensor, respectively.  Note that the static bulk equations guarantee that the latter obeys the sourced conservation equation
\be
\Dbar^\alpha \left\langle T_{\alpha\beta} \right\rangle = \left\langle \Ocal \right\rangle \Dbar_\beta \phib.
\ee

%
\section{Local Energy Bounds}
\label{sec:negativeE}
%

We now analyse the geometry governing static bulk solutions to deduce bounds on the local energy density. Following~\cite{HicWis15a,HicWis15b} we find it is convenient to write the bulk solutions in the ``optical frame'',
\be
\label{eq:optical}
ds^2 = \frac{\ell^2}{Z(x)^2} \left( - dt^2 + g_{ij}(x) dx^i dx^j \right),
\ee
where $(M, g_{ij})$ is a Riemannian 3-geometry which we term the ``optical geometry''. Then $Z(x)$ may be regarded as a function over~$M$.

We choose~$Z/\ell$ to be a defining function for the bulk conformal boundary, so that $Z \to 0$ at the bulk conformal boundary with $d Z \ne 0$. Then the optical geometry $(M, g_{ab})$ has a smooth boundary, $\partial M$, that corresponds to the spacetime asymptotic region. In particular our defining function~$Z/\ell$ is chosen so the corresponding representative of the conformal class of the boundary is the ultrastatic metric~\eqref{eq:boundary}, whose spatial metric $\gb_{ab}$ is given by the metric induced from $g_{ij}$ on $\partial M$.

It is again convenient to define a generalisation of the optical Ricci tensor to include the bulk scalar:
\begin{align}
L_{ij} = R_{ij} - \left(D_i \phi\right)\left( D_j \phi\right).
\end{align}
Then the static bulk equations, including the massless scalar, can be written covariantly in the optical geometry as
\begin{subequations}
\label{eqs:bulkOp}
\begin{align}
L_{ij} & = - \frac{2}{Z} D_i D_j Z, \\
L  & = \frac{6}{Z^2} \left( 1 - \left( D_i Z \right)^2 \right), \\
D^2 \phi & = \frac{2}{Z} (D^i Z)( D_i \phi). \label{eq:D2phi}
\end{align}
\end{subequations}

Let us now consider the behaviour of the optical geometry near the spacetime conformal boundary and near a horizon.  Near the spacetime boundary, we may read off the metric components~$g_{zz}$,~$g_{\alpha\beta}$ and defining function~$Z$ of the optical geometry from the FG expansion~\eqref{eqs:FG}.  More relevant to our purposes will be the expansions of~$D^2 \phi$,
\begin{align}
\label{eq:phiexpansion}
D^2 \phi & = 2 \Dbar^2 \phib + O(z),
\end{align}
as well as of the trace~$L \equiv g^{ij} L_{ij}$,
\be
\label{eq:Lexpansion}
L = 3 \Lbar - \frac{6\rhob}{c} \, z + O(z^2).
\ee
Thus the boundary quantity $\Lbar$ is given (up to a factor of three) by the bulk quantity $L$ restricted to the boundary:
\begin{align}
\label{eq:LpartialM}
L|_{\partial M} = 3 \Lbar.
\end{align}
In addition the normal gradient of $L$ at the boundary of the optical geometry determines the local energy density as
\be
\label{eq:rhofromL}
\left. \partial_n L \right|_{\partial M} = \frac{6}{c} \, \rhob
\ee
where $\partial_n$ is the outer-directed unit normal vector (in the optical geometry) to the boundary $\partial M$, given by $\partial_n = - \partial_z$ at $z =0$. 

To study the behaviour of the optical geometry near a horizon, we first note that a smooth finite temperature bulk horizon with surface gravity $\kappa$ (with respect to $(\partial_t)^a$) corresponds to an asymptotic region in the optical metric where $Z \to \infty$.  In fact, it is a conformal boundary.  Translating our previous expansion near the horizon from equations~\eqref{eqs:horizon}, we find the behaviour
\begin{subequations}
\bea
\frac{Z}{\ell} & = \frac{1}{\kappa\, r} \left( 1 + \frac{1}{12}  \Lcal r^2 + O(r^4) \right), \\
g_{rr} & = \frac{1}{\kappa^2 r^2}  \left( 1 + \frac{1}{6} \Lcal r^2  + O(r^4) \right),
\eea
\begin{multline}
g_{\alpha\beta} = \frac{1}{\kappa^2 r^2}  \left[ \gcal_{\alpha\beta} + \frac{1}{2} \left( \Lcal_{\alpha\beta} + \left( \frac{1}{3} \Lcal + \frac{3}{\ell^2} \right) \gcal_{\alpha\beta} \right) r^2 \right. \\ \left. \phantom{\left( \frac{1}{3} \Lcal + \frac{3}{\ell^2} \right)} + O(r^4) \right],
\end{multline}
\end{subequations}
where the scalar is still as given in \eqref{eq:horizonscalar}.  We see the spatial 2-geometry of the horizon $\gcal_{\alpha\beta}$ determines the conformal class of the conformal boundary of the optical metric.

Using this expansion we find the quantity~$L$ has the behaviour
\begin{align}
L & = - 6 \kappa^2 + \kappa^2 \left( 3 \Lcal + \frac{6}{\ell^2} \right) r^2 + O(r^4)
\end{align}
near the horizon.  Note then that at a horizon,~$L$ has the nice properties that it is constant, negative and simply determined by the temperature:
\begin{align}
\label{eq:horizonkappa}
\lim_{r\to 0} L & = - 6 \kappa^2.
\end{align}
Then the normal derivative of $L$ at the horizon conformal boundary, weighted by the local volume element, takes the form
\begin{align}
\label{eq:horizondnL}
\lim_{r\to 0} \left( \sqrt{g}\, \partial_n L \right) & = - 6 \kappa \sqrt{\gcal} \left( \Lcal + \frac{2}{\ell^2} \right), 
\end{align}
where $\partial_n$ is the outwards-pointing unit normal (in the optical geometry), given as $(\partial_n)^a = -(\kappa r)^{-1} (\partial_r)^a$ as $r \to 0$.  We pause here to note that when the scalar field is constant, the right-hand side of~\eqref{eq:horizondnL} is a combination of the Euler density and volume element of the horizon; we will discuss this observation more later.

The behaviour of the optical geometry, and in particular the quantity $\partial_n L$ both near the boundary and any horizon ends, allows us to derive a bound on the local energy density.  The key relation is
\begin{align}
\label{eq:D2Lneg}
D^2 L = - 3 \widetilde{L}_{ij} \widetilde{L}^{ij} - \frac{12}{Z^2} \left( (D^i Z)(D_i \phi) \right)^2,
\end{align}
where we have defined the traceless part of $L_{ij}$ as
\begin{align}
\widetilde{L}_{ij} \equiv L_{ij} - \frac{1}{3} g_{ij} L.
\end{align}
It is a straightforward computation to confirm that~\eqref{eq:D2Lneg} follows from the intermediate result (obtained using the Bianchi identity)
\begin{align}
\label{eq:Bianchi}
D_i L = \frac{6}{Z} (D^j Z) \widetilde{L}_{ij}.
\end{align}

The important consequence of~\eqref{eq:D2Lneg} is that
\be
\label{eq:lapL}
D^2 L \le 0.
\ee
This result is remarkably simple and will prove crucial in the derivation of Result~1, but a deeper understanding of its physical origin eludes us.  Understanding it better may be useful for formulating bounds in other contexts.

Let us briefly note when the inequality~\eqref{eq:lapL} may be saturated.  We see from~\eqref{eq:D2Lneg} that vanishing~$D^2 L$ requires both~$\widetilde{L}_{ij}$ and~$(D^i Z)(D_i \phi)$ to vanish; via~\eqref{eq:D2phi} and~\eqref{eq:Bianchi}, these imply that
\begin{subequations}
\bea
D^2 \phi & = 0, \\
L & = \mathrm{const}.
\eea
\end{subequations}
Via the expansions of $D^2 \phi$ and $L$ near the boundary (equations~\eqref{eq:phiexpansion} and~\eqref{eq:Lexpansion}), these conditions in turn imply
\begin{subequations}
\begin{align}
\Dbar^2 \phib & = 0, \\
\Lbar & = \mathrm{const}.
\end{align}
\end{subequations}
Since the boundary is closed the first condition above implies $\phib$ is constant, in which case the second implies~$\Rbar$ is constant as well.  Hence equality everywhere in equation~\eqref{eq:lapL} is only possible if we choose a constant source $\phib$ and boundary metric $\gb_{\alpha\beta}$ with a constant curvature $\Rbar$ (which will be determined by the boundary topology by Gauss-Bonnet).  Any smooth deformation of either the boundary scalar source or metric implies \eqref{eq:lapL} becomes a strict inequality somewhere in~$M$.

%
\subsection{Simple Bound for Confining Vacua}
%

We may quickly derive a simple bound on the local energy density in the case that the optical metric has no boundary other than the one corresponding to the spacetime conformal boundary. In this case the bulk falls into our confining vacuum class (though we note it also may be interpreted as having finite temperature, the effect of which is invisible at the level of semiclassical bulk gravity).

Then $D^2 L \le 0$ implies a minimum principle for $L$.  Since in this case $M$ is a smooth bounded geometry we may apply the Hopf minimum principle~\cite{GilbargT} which implies that unless $L$ is constant, then the minimum of $L$ must occur at the boundary $\partial M$ of the optical geometry.  Furthermore, the outer normal gradient is strictly negative there, so $\partial_n L < 0$.

Now since $L|_{\partial M} = 3 \Lbar$, the minimum of $L$ occurs precisely on the boundary where $\Lbar$ is minimised -- these are the points we labelled $p^\star$ in Section~\ref{sec:results} above.  Then for non-constant~$\Lbar$, the strictly negative outer normal gradient there determines that
\begin{align}
\label{eq:simplebound}
\rhob( p^\star ) < 0
\end{align}
directly from our boundary expansion \eqref{eq:rhofromL}.  Note that this is a strict inequality, and it holds even when there is more than one point~$p^\star$ at which~$\Lbar$ attains its global minimum; thus this result is slightly stronger than~\eqref{eq:result1cor}.

While this is a simple bound, it is highly non-trivial from a field theoretic perspective.  Simply knowing the spatial metric $\gb_{\alpha\beta}$ and scalar source $\phib$ (and hence $\Lbar$) for our CFT allows us to identify where the energy density is guaranteed to be negative, no matter what the behaviour of the expectation value of the stress tensor does elsewhere.

%
\subsection{Improved Bound}
%

We wish to improve on this simple bound in two regards. First, it requires a confining vacuum, so there are no other boundaries or ends to the optical geometry - i.e.~no horizons or singularities in the full bulk spacetime. However as we have emphasised, for toroidal boundaries null singularities are required already in the canonical case of  Poincar\'e-AdS suitably compactified. This is certainly expected to persist when the boundary metric is deformed and/or a scalar source is turned on. Second, we have a statement about the energy density only at the special points~$p^\star$.

Let us define a non-negative function $f(\alpha)$ which is monotonically decreasing and suitably smooth so that
\begin{align}
\label{eq:weightfn}
0 \le f(\alpha)  \; , \quad  f'(\alpha) \le 0.
\end{align}
Then we have
\begin{align}
D^i \left( f(L) D_i L \right) = f(L) D^2 L + f'(L) (D_i L)(D^i L),
\end{align}
which due to our constraints on the function $f$ and equation~\eqref{eq:lapL} yields
\begin{align}
\label{eq:boundeqn}
D^i \left( f(L) D_i L \right) \le 0.
\end{align}
Provided we have not chosen a pathological $f$ such that $f(L)$ vanishes everywhere in the bulk, then we may only have equality everywhere in the bound above if $L$ (and thus~$\Lbar$) is constant (and hence~$\Rbar$ and~$\phib$ are independently constant as well).

Integrating and using the divergence theorem we obtain the inequality
\begin{align}
\int_{\partial M} f(L) \star d L + \sum_i \lim_{\Hcal_i} \int f(L) \star d L \le 0,
\end{align}
where $\lim_{\Hcal_i}$ represents the contribution from asymptotic ends of the optical geometry corresponding to bulk horizons $\Hcal_i$, and we remind the reader that~$\star$ is the Hodge star.

Let us firstly examine the surface term at $\partial M$. Using our expansion~\eqref{eq:rhofromL} of~$L$ near this boundary, we obtain
\bea
\int_{\partial M} f(L) \star d L &= \int_{z =0} \sqrt{\gb} \, f(L) \partial_n L  \nonumber \\
		&= \frac{6}{c} \int_{\partial M} \sqrt{\gb} \, F(\Lbar) \rhob,
\eea
where $\partial_n$ is the outer unit normal, and we have defined for convenience
\begin{align}
F(\alpha) \equiv f(3 \alpha)
\end{align}
so that $F$ obviously shares the same properties~\eqref{eq:weightfn} as~$f$.  Hence this boundary term is simply an integral of the energy density weighted by the function $F$ evaluated on the boundary quantity $\Lbar$.

The surface term at the horizon can be evaluated remembering that $L$ is constant (and equal to $- 6 \kappa^2$) when restricted to a horizon. Using the expansion~\eqref{eq:horizondnL} near the horizon, we have
\begin{align}
\lim_{\Hcal_i} \int f(L) \star d L & =  f(- 6 \kappa^2) \lim_{r \to 0} \int \sqrt{\gcal} \, \partial_n L 
\nonumber \\
& = - 6 \kappa f(- 6 \kappa^2 ) \int_{\Hcal_i} \sqrt{\gcal} \left( \frac{2}{\ell^2} + \Lcal \right) ,
\end{align}
where $\partial_n$ is the outer unit normal to a constant $r$ surface.

Thus denoting $A_{\Hcal_i}$ as the area of the horizon component $\Hcal_i$ we obtain the inequality
\begin{align}
\int_{\partial M} \sqrt{\gb} F( \Lbar ) \rhob \le c \, \kappa F(- 2 \kappa^2 ) \sum_i \left( \frac{2}{\ell^2} A_{\Hcal_i} + \int_{\Hcal_i} \sqrt{\gcal} \Lcal \right) 
\end{align}
for any non-negative decreasing $F$.  Recall that~$\Lcal = \Rcal - (\D \varphi)^2$, so $\int_{\Hcal_i} \sqrt{\gcal} \Lcal$ consists of a contribution from the Euler characteristic of the horizon and a term involving the gradient of the scalar.
 
\begin{figure*}[t]
\centering
\subfigure[]{
\includegraphics[page=1,width=0.3\textwidth]{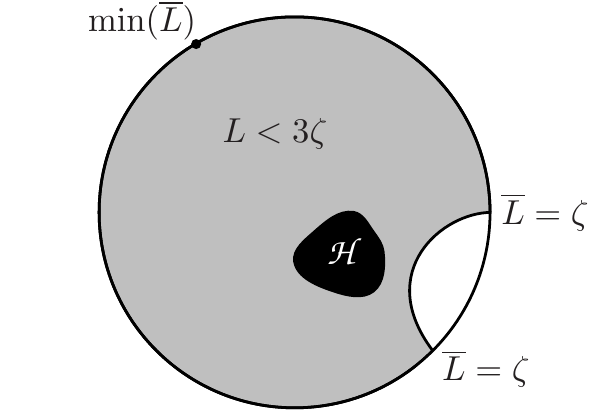}
\label{subfig:LH}
}
\subfigure[]{
\includegraphics[page=2,width=0.3\textwidth]{Figures-pics}
\label{subfig:LnoH}
}
\subfigure[]{
\includegraphics[page=3,width=0.3\textwidth]{Figures-pics}
\label{subfig:LHnoboundary}
}
\caption{The behaviour of~$F(L)$ for the step function~\eqref{eq:stepF};~$F(L)$ is unity in the grey regions and zero in the white regions.  The boundary region~$U_\zeta$ (defined in~\eqref{eq:Szeta}) corresponds to the portion of the boundary intersected by the grey region.  For simplicity, here we show a static time slice of the bulk in the case where the boundary has spherical topology and there is one bulk horizon~$\Hcal$ with surface gravity~$\kappa$.  \subref{subfig:LH}: the case where~$\zeta > \min(\Lbar)$ and~$\zeta > -2\kappa^2$, so~$E_\zeta$ contains a contribution from~$\Hcal$.  \subref{subfig:LnoH}: the case where~$\mathrm{min}(\Lbar) < \zeta < -2\kappa^2$, so~$E_\zeta$ contains no contribution from~$\Hcal$.  \subref{subfig:LHnoboundary}: the case where~$-2\kappa^2 < \zeta < \min(\Lbar)$, so the boundary region~$U_\zeta$ is trivial.  Then one obtains the bound~\eqref{eq:horizoncondition} relating the area and topology of the horizon to the scalar field on it.}
\label{fig:Fbehaviour}
\end{figure*}

In particular, suppose we choose $F$ to be a Heaviside function:
\be
\label{eq:stepF}
F(x) = \begin{cases} 1 & x <  \zeta \\ 0 & x \ge \zeta \end{cases}.
\ee
Then we obtain
\be
\label{eq:Ekappabounds}
E_\zeta \le 
\begin{cases}
c \, \kappa  \sum_i \left( \frac{2}{\ell^2} A_{\Hcal_i} +  \int_{\Hcal_i} \sqrt{\gcal} \Lcal \right) & - 2 \kappa^2 < \zeta \\
0 & \mathrm{otherwise}
\end{cases},
\ee
where the regions~$U_\zeta$ and corresponding energies~$E_\zeta$ were defined in~\eqref{eq:Szeta} and~\eqref{eq:Ezeta}, respectively.  Note that whether the contribution from the horizons is included depends on their temperature and the value of $\zeta$.  In Figure~\ref{subfig:LH}, we show a generic example (in the case of a single bulk horizon component) where~$\zeta > \min(\Lbar)$ and~$\zeta > -2\kappa^2$: in this case, the horizon contributes to~$E_\zeta$ (this will be the case for e.g.~small deformations of global AdS-Schwarzschild).  The other cases ($\min(\Lbar) < \zeta < -2\kappa^2$ and~$-2\kappa^2 < \zeta < \min(\Lbar)$) are shown in Figures~\ref{subfig:LnoH} and~\ref{subfig:LHnoboundary}.

We can rewrite~\eqref{eq:Ekappabounds} in terms of the temperature $T = \kappa / 2 \pi$ and entropies $S_{\Hcal_i} = 4 \pi c A_{\Hcal_i}/\ell^2$, the latter of which sum to give the total entropy $S = \sum_i S_{\Hcal_i}$.  We thus obtain a one-parameter set of bounds parameterised by~$\zeta$:
\be
\label{eq:bound1}
E_\zeta \le \begin{cases}
T  S + 2 \pi  c T \sum_i \int_{\Hcal_i} \sqrt{\gcal} \Lcal & - 2 \kappa^2 < r
\\
0 & \mathrm{otherwise}
\end{cases},
\ee
with equality possible only for constant boundary~$\Rbar$ and~$\phib$.

We have now derived a bound on the energy associated to the compact regions $U_\zeta$ on the boundary.  If~$\Lbar$ has an isolated global minimum at~$p^\star$, then when the bulk contains no horizons we can reproduce a slightly weaker version of~\eqref{eq:simplebound}: we simply take the limit~$\zeta \to \min(\Lbar)$ and obtain~$\rhob(p^\star) \leq 0$.  More generally, however, our bound allows for bulk horizons and applies not just to the energy density at this point.
 
We now make some remarks.  If there are no bulk horizons then we obtain simply $E_\zeta \le 0$ for any region $U_\zeta$.  However, we emphasise that in Section~\ref{sec:IMCF} we will show that it is not necessarily the case that $\rhob$ will be everywhere non-positive.

If there are bulk horizons, but $\min(\Lbar) < - 2 \kappa^2$, then we can choose $\zeta < - 2 \kappa^2$, as shown in Figure~\ref{subfig:LnoH}.  In such a case, there are non-trivial regions $U_\zeta$ where $E_\zeta \le 0$. 

If the boundary satisfies $\min(\Lbar) > - 2 \kappa^2$ (as in, for instance, global or planar AdS-Schwarzschild), then we can always choose a $\zeta$ such that the region~$U_\zeta$ is trivial and so the only contribution comes from the horizons; this is illustrated in Figure~\ref{subfig:LHnoboundary}. 
Then the bound implies the condition
\begin{align}
& \min(\Lbar) > - 2 \kappa^2 \quad \implies \nonumber \\
& \qquad  \sum_i \int_{\Hcal_i} \sqrt{\gcal} \left( \D \varphi\right)^2 <  \sum_i \left( \frac{2}{\ell^2} A_{\Hcal_i} + 4 \pi \chi_{\Hcal_i} \right),
\end{align}
with $\chi_{\Hcal_i}$ the Euler characteristic\footnote{Recall that for closed, orientable surfaces like~$\Hcal$, the Euler characteristic is related to the genus~$\mathfrak{g}_\Hcal$ as~$\chi_\Hcal = 2(1-\mathfrak{g}_\Hcal)$.} of the horizon component $\Hcal_i$, and we note that since $L$ is not constant over $M$ (since~$\Lbar \neq -2\kappa^2$) the inequality above is strict.  
In fact, this bound holds horizon component-by-component.  To see this, first note that~\eqref{eq:lapL} implies that $L$ has no minima in the interior of $M$.  Then assuming $\min(\Lbar) \ge - 2 \kappa^2$, from~\eqref{eq:LpartialM} we have~$\min(L |_{\partial M}) \ge - 6 \kappa^2$, and so~$L$ must obtain its minimum on the bulk horizons (since from~\eqref{eq:horizonkappa}, $L |_{\mathcal H_i} = - 6 \kappa^2$).  Hence the outer normal on the horizons must be non-positive,~$\partial_n L |_{\mathcal H_i} \le 0$, implying from~\eqref{eq:horizondnL} that~$\left. \left( \Lcal + 2/\ell^2 \right)\right|_{\mathcal H_i} \ge 0$ on all horizons.  Integrating this condition over just one horizon component $\mathcal H_i$ we thus conclude that
\begin{align}
\label{eq:horizoncondition}
& \min(\Lbar) \ge - 2 \kappa^2 \quad \implies \nonumber \\
& \qquad  \int_{\Hcal_i} \sqrt{\gcal} \left( \D \varphi\right)^2 \le  \left( \frac{2}{\ell^2} A_{\Hcal_i} + 4 \pi \chi_{\Hcal_i} \right).
\end{align}
Without a scalar source this bound is trivial unless the horizons have negative Euler characteristic (genus greater than one), in which case it provides a lower bound on the area, and hence entropy, associated to a black hole state. With a non-trivial boundary scalar source, the bound is non-trivial for any horizon topologies and can be viewed as bounding the magnitude of scalar variations on the horizons in terms of their area.

Now let us conclude this section by considering the implication of our bound for a (zero temperature) vacuum state.  First, consider a confined vacuum state (in which the bulk contains no boundaries besides the AdS boundary); then from~\eqref{eq:bound1} we immediately find that~$E_\zeta \leq 0$ for all~$\zeta$.  Next, consider a deconfined or degenerate vacuum.  Then provided that~$\int_{\Hcal_i} \sqrt{\gcal} (\D \varphi )^2$ is finite in the limit~$T \to 0$ (which we would require for a regular horizon), we have that $T S \to 0$ and $T \int \sqrt{\gcal} \Lcal \to 0$.  Thus again from~\eqref{eq:bound1} we find~$E_\zeta \leq 0$ for all~$\zeta$.  We have therefore obtained Result~1 (in fact, a slightly stronger version: Result~1 holds in the deconfined and degenerate vacua even if we remove the constraint requiring the topology of the bulk horizon to match that of the boundary).

%
\section{Inverse Mean Curvature Flow}
\label{sec:IMCF}
%

Having used simple Riemannian methods to consider bounds on the local energy density, we now consider the use of the inverse mean curvature (IMC) flow.  This flow was originally introduced by Geroch~\cite{Ger73} and extended by Jang and Wald~\cite{JanWal77} as a way of deriving Penrose inequalities in flat space (that is, a lower bound on the ADM mass of an asymptotically flat spacetime in terms of the area of any horizons), and was made rigorous in~\cite{HuiIlm01}.  Interestingly, the flow itself was generalised to an asymptotically hyperbolic setting by Lee and Neves~\cite{LeeNev15} (though its use in proving Penrose-like inequalities is subject to limitations, as shown in~\cite{Nev10}).

For the benefit of the reader we will review the construction of the flow, though we refer to~\cite{HuiIlm01,LeeNev15} for details and a rigorous construction and to~\cite{Bra02} for an introduction and overview.  In order to more closely connect to the results of the previous section, we will initially continue to restrict to a static bulk spacetime.  However, as we will discuss briefly later, the IMC flow methods reviewed here may be applied in a much more general (i.e.~dynamical) context.

We will  require the (massless) bulk scalar to be constant. This follows necessarily for a static bulk if there is a constant boundary source for the scalar: then since $\phi$ is harmonic in the bulk and has vanishing normal derivative at any horizons, it must be constant everywhere.  Thus for now we consider only pure gravity in the bulk - i.e.~the universal sector of AdS/CFT.

\subsection{Review of IMC Flow}

Consider a static time slice~$\Sigma$ of the bulk.  Using the Gauss-Codazzi equations, the staticity of this slice implies that its Ricci scalar is
\be
\label{eq:RG}
^\Sigma \! R = -\frac{6}{\ell^2}.
\ee
Let us foliate this slice by a one-parameter family of compact surfaces~$I_s$.  We construct this foliation by requiring that flowing along~$s$ (the ``flow time'') moves the surfaces along their normals with speed equal to their inverse mean curvature; that is, defining~$n^a$ to be the unit normal vector field to the~$I_s$ and~$^I \! K$ to be their mean curvature, we require
\be
\pounds_n s = \, ^I \! K.
\ee
This implies that the rate of change of a quantity with respect to~$s$ can be computed by a Lie derivative along~$^I \! K^{-1} n^a$; we will denote such derivatives with dots.  Thus denoting the induced metric on the surfaces~$I_s$ as~$\sigma_{ab}$, we have that the rate of change of~$\sigma_{ab}$ with respect to~$s$ is
\be
\dot{\sigma}_{ab} \equiv \pounds_{n/\, ^I \! K } \sigma_{ab} = \frac{1}{^I \! K} \pounds_n \sigma_{ab} = \frac{2}{^I \! K} \, ^I \! K_{ab},
\ee
where~$^I \! K_{ab}$ is the extrinsic curvature of the~$I_s$ in~$\Sigma$.

We now outline monotonicity results for the IMC flow.  First, note that the area of the surfaces~$I_s$ obeys
\be
\label{eq:Adot}
\dot{A}_I = \frac{1}{2} \int_I \sqrt{\sigma} \, \dot{\sigma}_{ab} \sigma^{ab} =  A_I,
\ee
so the area grows exponentially in flow time,~$A_I \propto e^s$.

Next, define the Hawking mass \cite{Hawking:1968qt} of the surfaces~$I_s$ to be
\bea
\label{eq:HawkingM}
m_H[I] & =  \sqrt{A_I} \, \int_I \sqrt{\sigma} \left( 2 \, ^I \! R - \, ^I \! K^2 + \frac{4}{\ell^2} \right), \nonumber \\
& = \sqrt{A_I}\left( 8 \pi \chi_{I} + \frac{4}{\ell^2} A_I -  \int_I \sqrt{\sigma} \, ^I \! K^2 \right),
\eea
where~$^I \! R$ is the Ricci scalar of~$I$ and we used the Gauss-Bonnet theorem to replace its integral with the Euler characteristic~$\chi_I$.  To obtain the behaviour of the Hawking mass along the flow, we first compute
\begin{multline}
\label{eq:Kdot}
^I \! \dot{K} = \frac{1}{2\, ^I \! K} \Bigg[ - \, ^\Sigma \! R + \, ^I \! R - \, ^I \! K^2 - \, ^I \! K_{ab} \, ^I \! K^{ab}\\ + 2 \, ^I \! K \, ^I \! D^a \left(\frac{^I \! D_a \, ^I \! K}{^I \! K^2}\right)\Bigg].
\end{multline}
Assuming the topology of the surfaces doesn't change,~$\chi_I$ is constant along the flow.  Then combining~\eqref{eq:Kdot} with~\eqref{eq:RG} and using~\eqref{eq:Adot}, we find that the Hawking mass increases monotonically along the flow:
\bea
\dot{m}_H &= \sqrt{A_I} \int_I \sqrt{\sigma}  \left(\left| \, ^I \! K_{ab} - \frac{1}{2} \, ^I \! K \sigma_{ab} \right|^2 + 2|^I \! D_a \ln \, ^I \! K|^2\right)\nonumber \\
		&\ge 0.
\eea
Note that if the IMC flow surfaces are homogeneous so that~$^I K_{ab} = \, ^I K \sigma_{ab}/2$ (as they will be in the case of e.g.~spherical symmetry), then the inequality above is saturated, and the Hawking mass is constant along the flow.

This monotonicity can be used to bound the asymptotic~($s \to \infty$) behaviour of the Hawking mass in terms of its value on some initial surface.  To that end, consider starting the IMC flow at the bifurcation surface~$\Hcal$ of a Killing horizon (for now, we furthermore assume that there is at most one such surface).  Since~$\Hcal$ is a minimal surface ($K = 0$), its Hawking mass is
\be
\label{eq:IMChoriz}
m_H[\Hcal] = \sqrt{A_{\Hcal}}  \left( 8 \pi \chi_{\Hcal} + \frac{4}{\ell^2} A_{\Hcal} \right).
\ee
Alternatively, in the absence of any Killing horizons, the flow may be started on the surface of a small geodesic ball centred on a point, in which case~$m_H$ vanishes in the limit the ball is taken very small.  In both cases these surfaces provide good initial conditions for the IMC flow at some finite starting flow time, say $s = 0$.

In asymptotically flat space, the asymptotic value of~$m_H$ is precisely the ADM mass, which by monotonicity of~$m_H$ is bounded from below by~\eqref{eq:IMChoriz} (with~$\ell \to \infty$): this yields a Penrose inequality.  In the asymptotically hyperbolic case, it was suggested in~\cite{Gib98} that the asymptotic value of~$m_H$ should again be the total mass of the spacetime, and thus still yield a Penrose inequality.  Interestingly,~\cite{Nev10} showed that this is not the case: an asymptotically hyperbolic space generally does not have sufficient convergence properties to guarantee that the IMC flow asymptotes to its total mass.

\begin{figure}[t]
\centering
\includegraphics[page=4,width=0.25\textwidth]{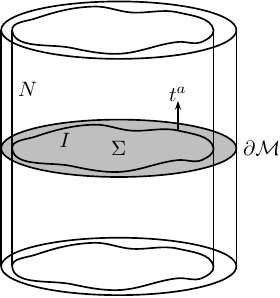}
\caption{Surfaces used to study the asymptotics of the IMC flow.  The flow surfaces foliate the asymptotic region of a bulk static time slice~$\Sigma$ (shaded grey); note that here we show just one surface~$I$ for simplicity.  We foliate the asymptotic region of the full bulk spacetime with timelike hypersurfaces~$N_s$ which intersect~$\Sigma$ on the flow surfaces~$I_s$.  The surfaces~$N_s$ are adapted to the time-translation isometry of the spacetime, so that the unit timelike vector field~$t^a$ normal to~$\Sigma$ is tangent to the~$N_s$.}
\label{fig:IMCasymptotics}
\end{figure}

To illustrate why this is the case, let us study the asymptotics of the IMC flow in more detail.  Assume that the flow surfaces~$I_s$ eventually approach the asymptotic boundary~$\partial M$ (in the precise way described in Appendix~\ref{app:IMCF}); then in any FG coordinate system, as $s \to \infty$ the flow asymptotes to
\be
z \sim f(x) e^{-s/2}
\ee
for some smooth positive definite function $f(x)$ of the boundary coordinates $x^\alpha$. Thus a flow that asymptotes to the AdS boundary defines a ``flow frame'', whose defining function we may take to be~$\widetilde{\Omega} = e^{-s/2}$ near the conformal boundary, such that the flow asymptotes to constant coordinate surfaces $\tilde{z} \sim e^{-s/2}$ in the associated FG coordinates.

The important point is that flows starting with different initial conditions in the interior of the bulk will generally have asymptotics that yield different flow frames. This is nicely illustrated with a simple example. Take (a time slice of) global AdS and consider flows that start at a point in the interior. In usual global AdS coordinates, a flow starting at the origin of course respects the spherical symmetry of the coordinates. As $s \to \infty$ it asymptotes to the boundary on constant radial coordinate slices, picking out the flow frame which is the natural one where the boundary metric is the ultrastatic product of time with a round sphere. However, we show in Appendix~\ref{app:AdSIMCF} that starting the flow ``off centre'' results in a flow frame which is not the ultrastatic one. 

Given that there is no preferred frame to which flows asymptote, then the question is whether the asymptotic Hawking mass is a conformal invariant or not, i.e.~if it is independent of the frame. As we now show, this is not the case, and while the Hawking mass does indeed depend on the energy density $\bar{\rho}$ in the ultrastatic conformal frame, it also depends on the data specifying the frame to which the flow asymptotes. 

In the asymptotic region let us introduce a foliation of timelike hypersurfaces~$N_s$ whose intersections with the bulk time slice~$\Sigma$ are just the IMC flow surfaces~$I_s$, as shown in Figure~\ref{fig:IMCasymptotics}.  We take these hypersurfaces to be adapted to the time-translation isometry of the spacetime, i.e.~$(\partial_t)^a$ is tangent to the~$N_s$.  Defining~$\gamma_{ab}$ and~$G_{ab}$ to be the induced metric and Einstein tensor of the~$N_s$, their extrinsic curvature~$^N \! K_{ab}$ is related to the Balasubramian-Kraus boundary stress tensor~\cite{BalKra99} as\footnote{Some signs here differ from the original reference~\cite{BalKra99} due to differing curvature conventions.}
\be
\label{eq:BalaKraus}
\T_{ab} = \frac{2c}{\ell^2}\left[- \, ^N \! K_{ab} + \, ^N \! K \gamma_{ab} - \frac{2}{\ell} \gamma_{ab} + \ell G_{ab}\right].
\ee
We remind the reader that~$\T_{ab}$ is defined as a bulk object, and in particular makes no use of a choice of conformal frame.  However, once a conformal frame~$\Omega$ is chosen, the boundary stress tensor in this frame is given by~$\left\langle T_{ab} \right\rangle = \lim_{N \to \partial M} \T_{ab}/\Omega$.

The asymptotic value of the Hawking mass can be expressed as an integral of~$\T_{ab}$.  To do so, first we note that the extrinsic curvature~$^I \! K_{ab}$ of~$I$ in~$\Sigma$ is just the projection of~$^N \! K_{ab}$ onto~$I$.  Then the mean curvatures are related by
\be
^I \! K = \, ^N \! K + \, ^N \! K_{ab}t^a t^b,
\ee
where~$t^a = (\partial_t)^a/\sqrt{-\|\partial_t\|^2}$ is the unit timelike normal vector field to~$\Sigma$ (and which by construction is therefore tangent to~$N$).  We may then use~\eqref{eq:BalaKraus} to express~$^I \! K$ in terms of boundary quantities:
\be
\label{eq:Ksigma}
^I \! K = \frac{2}{\ell} - \frac{\ell^2}{2c} \T_{ab} t^a t^b + \ell G_{ab} t^a t^b.
\ee

Next, note that since the time slice~$\Sigma$ is static, its extrinsic curvature vanishes.  The Gauss-Codazzi equations then yield
\be
\label{eq:RI}
^I \! R = 2G_{ab} t^a t^b,
\ee
so that plugging this and~\eqref{eq:Ksigma} into~\eqref{eq:HawkingM}, we obtain that asymptotically,
\be
m_H = \frac{2\ell}{c} \sqrt{A_I} \int_I \sqrt{\sigma} \left(\T_{ab} t^a t^b + \Ocal(\Omega^4)\right),
\ee
where we have used the fact that near the boundary and for arbitrary conformal factor~$\Omega$,~$G_{ab} t^a t^b \sim \Ocal(\Omega^2)$.  

Neither of the two integrals that appear in~$m_H$ (recall that~$A_I$ is an integral over~$I$) can be written as conformal invariants of boundary data. This demonstrates our previous claim, that the asymptotic limit of the Hawking mass is not a conformal invariant of boundary quantities (and thus need not yield the total energy of the CFT). Hence the details of the specific flow will be important through the particular flow frame it picks out.

We may re-express the asymptotic behaviour of the Hawking mass in terms of boundary data in a particular frame.  Consider, for instance, working in the flow frame defined by~$\widetilde{\Omega} = e^{-s/2}$; then one finds
\be
m^\infty_H \equiv \lim_{s \to \infty} m_H[I_s] = \frac{2\ell}{c} \sqrt{\int_{\partial M} \sqrt{^{(\mathrm{f})} \! \gb}} \int_{\partial M} \sqrt{^{(\mathrm{f})} \! \gb} \, ^{(\mathrm{f})} \! \rhob,
\ee
where~$^{(\mathrm{f})} \! \gb_{ab}$ and~$^{(\mathrm{f})} \! \rhob$ are the spatial boundary metric and local energy density in the flow frame, respectively.  However, this frame will generically not be the same as the ultrastatic conformal frame with defining function $Z/\ell$; the function $\omega \equiv \lim_{\to \partial M} \left(\ell \widetilde{\Omega}/Z \right)$ describes the relation between these.  Note that $\omega$ is positive definite and will depend on the details of the flow, which in turn depend on its initial conditions and the bulk geometry.  Converting to this ultrastatic frame, we find
\be
\label{eq:defnI}
m^\infty_H = \frac{2\ell}{c} \sqrt{\int_{\partial M} \sqrt{\bar{g}} \, \omega^2} \int_{\partial M} \sqrt{\bar{g}} \, \frac{\rhob}{\omega}.
\ee

Now consider starting the IMC flow on some surface~$I_0$ at~$s = 0$. Then let us assume a smooth flow exists that asymptotes to the boundary. Note that a necessary condition for the flow to be smooth is that the topology of the surface $I_0$ is the same as that of the boundary~$\partial M$. Then monotonicity of Hawking mass under the IMC flow implies the highly non-trivial result
\be
\label{eq:monotonicM}
\lim_{s \to \infty} m_H[I_s] = m^\infty_H \ge  m_H[I_0].
\ee

\subsection{A Rigorous Bound}
\label{subsec:LeeNev}

This result~\eqref{eq:monotonicM} assumed the existence of a smooth IMC flow from an initial surface~$I_0$ to the asymptotic boundary. Certainly for static bulks with homogeneous spatial symmetry one would expect such flows to exist. The question in our context is: does such a smooth IMC flow exist for a generic (static) bulk?  The answer is no: the surfaces~$I_s$ can e.g.~cusp or encounter ``obstacles'' like black hole horizons.  In such cases, the above analysis breaks down.  However,~\cite{HuiIlm01,LeeNev15} rigorously showed that monotonicity of the Hawking mass is still preserved by a so-called weak solution to the IMC flow.  Roughly speaking, instead of thinking of~$s$ as a parameter along the family of surfaces~$I_s$, one takes~$s$ to be a scalar on the time slice~$\Sigma$ and defines the surfaces~$I_s$ as level sets of~$s$ whose asymptotic behaviour is that described in Appendix~\ref{app:IMCF}.  These level sets are then allowed to ``jump'' discontinuously to avoid singularities of the flow while still preserving monotonicity of the Hawking mass. However in order to obtain the bound \eqref{eq:monotonicM} for this weak flow we still require the topology of the starting surface and the boundary $\partial M$ to be the same \cite{LeeNev15}.

An important technical point is that~\cite{LeeNev15} assumed that the boundary~$\partial M$ was a constant-curvature space (i.e.~a round sphere, flat torus, etc...). In the static AdS/CFT setting, for constant-curvature boundary we know the exact bulk solution (AdS-Schwarzschild), whereas here we are precisely interested in the case that the boundary metric is non-trivially curved so that the bulk solution is unknown.  To fill in this small gap, in Appendix~\ref{app:IMCF} we generalise Lemma~3.1 of~\cite{LeeNev15}, which is where the details of the asymptotics appear, in order to use the Weak Existence Theorem~3.1 of~\cite{HuiIlm01} to guarantee that the weak IMC flow asymptotes to the asymptotic boundary~$\partial M$.  Then the proof presented in~\cite{LeeNev15} applies to our time slice~$\Sigma$ as well.

Thus the monotonicity result~\eqref{eq:monotonicM} above can be made mathematically rigorous and holds even when a smooth IMC flow does not exist.  Moreover, we may apply it even when there exists more than one Killing horizon, in which case we may start the flow at any bifurcation surface~$\Hcal$.  We refer to~\cite{LeeNev15} for details of this construction.  

Although not immediately relevant for the purpose of this paper, we should also pause to note that the IMC flow can be applied more generally than just to static time slices of static spacetimes.  Indeed, the analysis presented above will go through unchanged if~$\Sigma$ is only taken to be a moment of time symmetry (rather than a static time slice).  In particular, equations~\eqref{eq:RG} and~\eqref{eq:RI} will be unchanged, while~$\Hcal$ must be taken to be an apparent horizon rather than the bifurcation surface of a Killing horizon.  In fact, all that is really needed for the monotonicity result to go through is for~$^\Sigma R \geq -6/\ell^2$; this provides considerable freedom in both the choice of time slicing and the matter content of the spacetime.

\subsection{Derivation of Result 2}

Despite the fact that the asymptotic behaviour of the Hawking mass is not a conformal invariant, we may nevertheless use its monotonicity to derive constraints on the local energy density of the CFT.  To that end, consider first the case where the boundary has spherical topology,~$\mathfrak{g} = 0$.  Starting the IMC flow at flow time $s = 0$ at some bulk point, we find~$m_H[I_0] = 0$.  Second, consider the case where the boundary has toroidal topology,~$\mathfrak{g} = 1$, and the bulk contains a (bifurcation surface of a) nonzero-temperature horizon~$\Hcal$ of the same topology.  Starting the flow on~$\Hcal$, we note that since~$\chi_\Hcal = 0$ we have~$m_H[I_0] = m_H[\Hcal] \geq 0$.  Thus in both these cases monotonicity of the IMC flow implies from~\eqref{eq:defnI} and~\eqref{eq:monotonicM} that
\be
\label{eq:posM}
m_H^\infty \ge 0 \; \implies \; \int_{\partial M} \sqrt{\bar{g}} \, \frac{\rhob}{\omega}  \ge 0.
\ee
Since $\omega > 0$, the above inequality immediately implies that~$\rhob$ cannot be everywhere negative - this is essentially the~$\mathfrak{g} = 0$ or~$1$ case of Corollary 1.2 of~\cite{LeeNev15}, suitably generalised to asymptotically locally AdS boundary conditions in which the boundary spatial metric is not of constant curvature.

Taking the zero-temperature limit, we find that for~$\mathfrak{g} = 0$ the above result continues to hold in all three of the confined, deconfined, and degenerate vacuum classes.  For~$\mathfrak{g} = 1$, the requirement that the bulk contain a horizon of toroidal topology restricts us to the deconfined and degenerate vacuum classes only.  In addition, given that Result~1 implies that~$\rhob$ must be strictly negative somewhere if the boundary is not of constant curvature, we conclude from~\eqref{eq:posM} that~$\rhob$ must in fact be positive elsewhere (though we do not learn where).

The situation is more subtle for~$\mathfrak{g} > 1$.  Again, assume that at nonzero temperature the bulk contains a horizon~$\Hcal$ with the same topology as the boundary; then we may start the IMC flow there and obtain~$m^\infty_H \ge m_H[\Hcal]$.  However, for~$\mathfrak{g} > 1$, we have~$\chi_\Hcal < 0$, and thus~$m_H[\Hcal]$ need not be positive; this prevents us from recovering the positivity result~\eqref{eq:posM}.  We return shortly to considering the~$\mathfrak{g} > 1$ finite temperature case, but for now focus on its zero-temperature limit and consider only the deconfined vacuum. Then we have~$A_\Hcal \to 0$ and thus from~\eqref{eq:IMChoriz} we find $m_H[\Hcal] \to 0$.  Then again we recover~\eqref{eq:posM}, and thus as for lower genus,~$\rhob$ must be non-negative somewhere (and strictly positive somewhere if the boundary does not have constant curvature).

Note an important difference between the result for~$\mathfrak{g} = 1$ and that for~$\mathfrak{g} > 1$.  For~$\mathfrak{g} = 1$, we expect a deconfined or degenerate vacuum to exist: in the case of the flat torus, the bulk is a quotient of the Poincar\'e patch of pure AdS, which as discussed in Section~\ref{sec:results} falls into the deconfined vacuum class.  Presumably, a (topology-preserving) perturbation of the boundary will result in a bulk which still ends on a null surface that is the zero-temperature limit of a regular horizon.  However, for~$\mathfrak{g} > 1$ it's not so clear what a deconfined vacuum would be: for constant curvature boundary the bulk may be filled in by the zero temperature limit of the hyperbolic Schwarzschild-AdS black hole, but again as discussed in Section~\ref{sec:results} this has finite area and therefore falls into the degenerate vacuum class.

At finite temperature, we can obtain additional nontrivial results when the boundary and bulk horizon have genus higher than a torus.  In this case,~$\chi_{\Hcal} < 0$, so we may have~$m^\infty_H < 0$.  Indeed, for hyperbolic black holes~\cite{Emp99},~$\rhob$ may be everywhere negative for sufficiently low temperature.  If we assume that~$\rhob$ is everywhere negative (as it will be for~e.g.~a sufficiently small perturbation of a negative-mass hyperbolic black hole), we may obtain a useful result as follows.  First, let us consider~$m^\infty_H  = m^\infty_H[\omega]$ to be a functional of ~$\omega$ which describes the frame chosen by a particular IMC flow (relative to the ultrastatic one).  Then by functionally differentiating~$m^\infty_H[\omega]$ with respect to~$\omega$, we find that~$m^\infty_H[\omega]$ is maximised when~$\omega_\mathrm{max} = | \rhob |^{1/3}$, which implies
\be
\label{eq:Omegamax}
\rhob < 0 \Rightarrow m^\infty_H \le m^\infty_H[\omega_\mathrm{max}] = - \frac{2\ell}{c} \left( \int_{\partial \mathcal{M}} \sqrt{\bar{g}} | \rhob |^{2/3} \right)^{3/2}.
\ee
One may wonder why such an extremisation can only be performed when~$\rhob < 0$ everywhere.  In fact,~$\omega = \pm \rhob^{1/3}$ always extremise~$m^\infty_H[\omega]$.  However, since~$\omega > 0$, these extrema are inadmissible if~$\rhob$ has indefinite sign, while if~$\rhob > 0$ everywhere, then these extrema are \textit{minima} of~$m^\infty_H[\omega]$.  We refer to~\cite{LeeNev15} for more details.

Combining~\eqref{eq:Omegamax} with the IMC weak flow, we thus obtain that if~$\rhob < 0$ everywhere,
\bea
\label{eq:generalIMC}
\sqrt{A_{\Hcal}}  \left( 8 \pi \chi_{\Hcal} + \frac{4}{\ell^2} A_{\Hcal} \right) \le - \frac{2 \ell}{c}\left( \int_{\partial \mathcal{M}} \sqrt{\bar{g}} | \rhob |^{2/3} \right)^{3/2}.
\eea
We note that for hyperbolic AdS-Schwarzschild the bound is saturated since the spatial slices at constant Schwarzschild coordinate are homogeneous spaces.  Hence for deformations of the boundary geometry that preserve $\rhob < 0$, we expect the above bound to hold as an inequality.

Now, suppose the bulk contains a horizon~$\Hcal$ with the same topology~$\mathfrak{g} > 1$ as the boundary.  Then for vanishing scalar our result~\eqref{eq:horizoncondition} becomes
\be
\label{eq:horizoncondition2}
2 \kappa^2 \ge -\min(\Rbar) \quad \implies
\quad 2 \pi \chi_{\Hcal} + \frac{1}{\ell^2} A_{\Hcal}  \ge 0.
\ee
We emphasise that for a boundary with $\mathfrak{g} > 1$, Gauss-Bonnet ensures $\min(\Rbar) < 0$, so this statement is non-trivial.  Interestingly, the same combination~$2\pi \chi_\Hcal + A_\Hcal/\ell^2$ appears in both~\eqref{eq:generalIMC} and~\eqref{eq:horizoncondition2}, though these two inequalities were obtained via very different approaches (indeed, the form of~\eqref{eq:horizoncondition2} can be traced back to the form~\eqref{eq:horizondnL} of the normal derivative~$\partial_n L$ on a horizon).  The reason for the appearance of this evidently universal term is still unclear to us.  However, we may exploit this feature by combining~\eqref{eq:generalIMC} and~\eqref{eq:horizoncondition2} to derive that $\rhob$ cannot be everywhere negative if the temperature $T = \kappa/2\pi$ is sufficiently high compared to $\min(\bar{R})$:
\begin{align}
T^2 \ge -\min(\Rbar)/(8 \pi^2) \; \implies \; \exists p \in \partial \mathcal M \; \mathrm{so} \; \rhob(p) \ge 0.
\end{align}
Interestingly this bound may be saturated for the case of a constant curvature hyperbolic boundary, in which case the bulk dual is a quotient of hyperbolic AdS-Schwarzschild.  For these solutions at low temperatures the energy density is homogeneous and negative, and for high temperatures it is homogeneous and positive, with the transition exactly when $T^2 = -\min(\Rbar)/(8 \pi^2)$.

We now collect these various results, the vacuum versions of which are also summarised in Result~2~\eqref{eq:result2}:
\begin{itemize}
\item $\mathfrak{g} = 0$: $\rhob \ge 0$ somewhere for finite temperature or any of the three classes of vacuum.
\item $\mathfrak{g} = 1$: $\rhob \ge 0$ somewhere in either a deconfined or degenerate vacuum.  Also~$\rhob \ge 0$ somewhere at finite temperature if there exists a bulk horizon with~$\mathfrak{g} = 1$.
\item $\mathfrak{g} > 1$: $\rhob \ge 0$ somewhere in a deconfined vacuum.  At finite temperature if there exists a bulk horizon with genus~$\mathfrak{g}$ then $\rhob \ge 0$ somewhere if the temperature obeys~$T^2 \ge -\min(\Rbar)/(8 \pi^2)$.
\end{itemize}

The above results are consistent with the fact that for the AdS-soliton, which is a confined vacuum with~$\mathfrak{g} = 1$, the energy density is everywhere strictly negative.  Adding a sufficiently small localised black hole to the AdS-soliton (the low energy limit of a plasmaball~\cite{Aharony:2005bm,Figueras:2014lka}) to give a finite temperature solution is expected to leave the energy density everywhere negative, consistent with the fact that the horizon has non-toroidal topology.  Now, consider growing the black hole to the point where it undergoes a topology changing transition \cite{Horowitz:2011cq} and becomes toroidal; to be consistent with our results, the energy density must become positive somewhere before (or at) this transition.  Thus the positivity properties of~$\rhob$ may be interpreted as constraints on the allowed topology of dual bulk horizons (or vacua).

\subsection{Some Further Remarks for Higher Genus}

We conclude our discussion of bounds derived from the IMC flow by combining the result~\eqref{eq:generalIMC} with our previous result in equation~\eqref{eq:bound1} to obtain a bound on the variation of~$\rhob$.  We will again assume that~$\rhob < 0$ everywhere, and moreover that the bulk contains \textit{precisely one} horizon~$\Hcal$ of topology~$\mathfrak{g} > 1$.  First, note that
\be
\left( \int_{\partial M} \sqrt{\bar{g}} | \rhob |^{2/3} \right)^{3/2} \geq A_\mathrm{CFT}^{3/2} |\rhob_\mathrm{max}|,
\ee
where~$A_\mathrm{CFT}$ is the area of~$\partial M$ in the ultrastatic conformal frame, and~$\rhob_\mathrm{max}$ is the maximum value of~$\rhob$ over~$\partial M$.  Combining with~\eqref{eq:generalIMC}, we thus reproduce Theorem~1.1 of~\cite{LeeNev15}, suitably generalised to non-constant curvature boundary metrics: if~$\rhob < 0$ everywhere, then
\be
\label{eq:ineq1}
\sqrt{A_\Hcal}  \left( 8 \pi \chi_\Hcal + \frac{4}{\ell^2} A_\Hcal \right) \leq \frac{2\ell}{c} A_\mathrm{CFT}^{3/2} \rhob_\mathrm{max}.
\ee

Next, recalling that the total CFT energy~$E$ is given by
\be
E = \int_{\partial M} \sqrt{\bar{g}} \, \rhob \geq A_\mathrm{CFT} \rhob_\mathrm{min},
\ee
then turning off the scalar in our earlier result \eqref{eq:bound1} and taking~$\zeta$ sufficiently large (or equivalently the result of~\cite{HicWis15b}) yields
\be
\label{eq:ineq2}
A_\mathrm{CFT} \rhob_\mathrm{min} \leq E \leq 2\pi c  T \left(4\pi \chi_\Hcal + \frac{2}{\ell^2} A_\Hcal\right),
\ee
where~$\Hcal$ is (the bifurcation surface of) a bulk Killing horizon.  Combining~\eqref{eq:ineq1} and~\eqref{eq:ineq2}, we find that if~$\rhob < 0$ everywhere,
\be
\label{eq:rhoineq}
\left(\frac{\rhob_\mathrm{max}}{\rhob_\mathrm{min}}\right)^2 \leq \frac{s}{16\pi^3 cT^2},
\ee
with~$s \equiv 4\pi c A_\Hcal/(\ell^2 A_\mathrm{CFT})$ the CFT entropy density and~$T$ its temperature.

Note that by assumption,~$\rhob_\mathrm{min} \leq \rhob_\mathrm{max} < 0$ (so the bound presumably only applies at sufficiently low temperatures), so that~$\rhob_\mathrm{max}/\rhob_\mathrm{min} \leq 1$.  It is therefore interesting to ask whether the bound~\eqref{eq:rhoineq} is ever nontrivial.  We can examine this question by again considering hyperbolic AdS-Schwarzschild.  In this case, the answer is marginal.  At low temperatures the energy density is indeed negative, but the right-hand side of~\eqref{eq:rhoineq} becomes arbitrarily large as~$T \to 0$ since~$s$ does not vanish in this limit; hence for sufficiently low temperature, the bound is trivially satisfied.  On the other hand, at high temperatures the right-hand side tends to~$s/(16 \pi^3 c T^2) \to 4/9 < 1$ but the energy density $\bar{\rho}$ is positive.  Only at the single critical temperature~$T = 1/(2\pi\ell)$ at which the energy density vanishes (and hence the bulk is locally pure AdS) the bound is saturated, so that $\rhob$ is still non-positive and the right-hand side is not greater than one.  If one deforms the boundary metric away from constant negative curvature, it is then unclear whether there will be a range of intermediate temperatures where~$s/(16\pi^3 cT^2) < 1$ and $\bar{\rho} < 0$ everywhere (so that~\eqref{eq:rhoineq} imposes a nontrivial constraint).

As a final note, let us reproduce the~$\mathfrak{g} > 1$ case of Corollary~1.2 of~\cite{LeeNev15} (generalised to non-constant curvature boundary).  It is straightforward to show that for~$A_\Hcal \geq 0$ and~$\chi_\Hcal < 0$,
\be
\sqrt{A_\Hcal}  \left( 8 \pi \chi_\Hcal + \frac{4}{\ell^2} A_\Hcal \right) \geq -\left(-\frac{8\pi\chi_\Hcal}{3}\right)^{3/2}\ell;
\ee
combining this result with~\eqref{eq:ineq1}, we obtain
\be
\rhob_\mathrm{max} \geq -32c\left(\frac{\pi(\mathfrak{g}-1)}{3A_\mathrm{CFT}}\right)^{3/2},
\ee
where we replaced~$\chi_\Hcal = 2(1-\mathfrak{g})$.

%
\section{Discussion}
\label{sec:disc}
%

We have obtained interesting local constraints on the energy density in $(2+1)$-dimensional holographic CFTs with an Einstein-scalar gravity bulk and deformations of both the boundary metric and scalar source.  These bounds derive from the geometric nature of the bulk description, and are general enough to also apply to zero temperature vacuum geometries which have ``good'' singularities.  The bound derived in Section~\ref{sec:negativeE} generalises the previous global result of~\cite{HicWis15b}, while that derived in Section~\ref{sec:IMCF} based on the IMC flow results of~\cite{LeeNev15} is qualitatively new.

Recall per the discussion in Section~\ref{sec:intro} that these bounds are interesting in light of the facts that~(i) the energy density (and more generally, the stress tensor) is a local operator that exists in any QFT, and~(ii) energy bounds even for free, minimally-coupled theories are very difficult to derive.  These latter bounds -- quantum energy inequalities -- place lower bounds on the energy density, typically integrated along some curve or spacetime region.  It is thus very interesting to note that some of our results here are quite opposite in spirit: Result~1 in fact places an \textit{upper} bound on the (integrated) energy density of a holographic CFT!  This unusual feature is tempered slightly by our Result~2, which instead serves to bound the energy density from below with limited constraints for~$\mathfrak{g} = 0$, and with constraints on the topology of bulk horizons for~$\mathfrak{g} \geq 1$.  Moreover, our results only apply to static states (though for Result~2, all we really required was a moment of time symmetry).  Nevertheless, it would be very interesting to understand if approaches similar to those used in this paper can produce bounds that hold in much more general settings.  We now list some of these potential generalisations.

\textbf{Higher Dimensions.}  The discussion presented here has been limited to~$(2+1)$ boundary dimensions.  Do our results generalise to other dimensions?  Unfortunately, a na\"ive extension fails.  The elliptic equations derived in Section~\ref{sec:negativeE} can of course be extended to higher dimensions; however, when a higher-dimensional analogue of~\eqref{eq:boundeqn} is integrated over the bulk optical geometry, one finds that additional terms in the expansion of~$L$ near the boundary cause any contributions from~$\partial M$ to diverge (for general boundary spatial geometry).  This divergence prevents the extraction of the energy density in a meaningful inequality.

Likewise, the IMC flow used in Section~\ref{sec:IMCF} can be generalised to higher dimensions.  However, recall that a crucial ingredient in proving the monotonicity of the Hawking mass was the Gauss-Bonnet theorem, used to replace the integral of the Ricci scalar~$^I \! R$ with the Euler characteristic~$\chi_I$ (which is constant along the flow).  This replacement is special to two-dimensional flow surfaces~$I$, and thus to our knowledge efforts to generalise monotonicity of the Hawking mass in higher dimensions have proved unsuccessful.

\textbf{Bounds on other one-point functions.}  The bounds we have derived on the energy density came from purely geometric bulk considerations.  Given that in Section~\ref{sec:negativeE} we allowed the presence of a scalar, it is natural to ask if similar geometric arguments can be used to obtain bounds on the one-point function~$\left\langle\Ocal\right\rangle$.  Indeed, it is possible to obtain a harmonic inequality for the scalar field: introducing a weighting function~\eqref{eq:weightfn}, we obtain from the scalar equation of motion~\eqref{eqs:bulkOp} that
\be
D^i \left( \frac{f(\phi)}{Z^2} D_i \phi \right) \le 0.
\ee
One might then hope that using the divergence theorem we would again obtain an inequality involving surface terms that are integrals of~$f D_i \phi/Z^2$.  Note, however, that~$D_i \phi/Z^2$ is singular at the boundary; indeed, a more careful analysis shows that any contribution to such surface terms from~$\partial M$ will diverge for generic boundary source.  Hence we cannot straightforwardly apply our method to constrain the one-point function~$\left\langle\Ocal\right\rangle$.

One might hope to circumvent this issue by instead considering a relevant scalar so that the bulk field~$\phi$ has a faster falloff near the boundary.  Adding a mass $m^2$ (which for a relevant deformation must be negative but above the BF bound~\cite{Breitenlohner:1982bm,Breitenlohner:1982jf}), one finds the scalar obeys the equation
\begin{align}
\label{eq:bulkOpscalar}
D^2 \phi & = \frac{2}{Z} (D^i Z)(D_i \phi) + \frac{m^2}{Z^2} \phi.
\end{align}
Then performing the same manipulations we obtain
\begin{align}
D^i \left( \frac{f(\phi)}{Z^2} D_i \phi \right) \le \frac{m^2}{Z^4} \phi f(\phi).
\end{align}
If~$m^2 < 0$, the right-hand side prevents us from obtaining a useful extremum principle on~$\phi$, so this modification is insufficient to allow us to obtain bounds on~$\left\langle \Ocal \right\rangle$ using the same techniques exploited in Section~\ref{sec:negativeE}.

\textbf{Additional bulk matter.}  Of course, along these lines a natural question is whether our analyses may be extended to derive constraints on the local energy density in the presence of other types of matter.  We note that we have not managed to generalise the bounds of Section~\ref{sec:negativeE} for the simplest generalisation of a massive bulk scalar.  The IMC flow, on the other hand, can be generalised to include matter; as noted in Section~\ref{subsec:LeeNev}, there is considerable freedom to modify the matter content of the bulk spacetime while still preserving monotonicity of the Hawking mass under IMC flow.  However, if one includes a massless bulk scalar field with nontrivial boundary source as we have considered in this paper, the asymptotic Hawking mass diverges, so its monotonicity can no longer be used to compare the CFT local energy density to bulk geometric objects.

\textbf{Dynamical states.}  Here we have only considered static CFT states, resulting in a static bulk geometry.  This staticity was crucial in deriving the elliptic equation~\eqref{eq:lapL} from which we derived Result~1.  However, as alluded to in Section~\ref{sec:IMCF}, the IMC flow can be applied to very general time slices: the only constraint required to preserve monotonicity of the Hawking mass is that~$^\Sigma \! R \geq -6/\ell^2$.  Thus all the results presented in Section~\ref{sec:IMCF} generalise as they are to a slice of time symmetry, and in fact generalise to much more general time slices of a general dynamical bulk.  Thus it is quite likely that bounds similar to Result~2 may be found on the local energy density in an arbitrary dynamical CFT state.

\textbf{Entanglement entropy.}  To derive the bounds in Section~\ref{sec:IMCF}, the IMC flow was started on the bifurcation surface of a Killing horizon; because such a surface is minimal, its Hawking mass can be expressed purely in terms of its intrinsic geometry (i.e.~its area and genus).  But recall that minimal surfaces play a special role in AdS/CFT: by the Ryu-Takayanagi (RT) proposal~\cite{RyuTak06,LewMal13}, the area of a minimal surface anchored to the AdS boundary computes the entanglement entropy of a spatial region of the CFT.  It would be very interesting to understand if starting the IMC flow on such a surface (which we emphasise is typically not compact) can yield any interesting constraints on CFT entanglement entropy.
 
We leave a more detailed exploration of these directions for future work.

%
\section*{Acknowledgements}
%

It is a pleasure to thank Jerome Gauntlett, William Kelly, Donald Marolf, and Antony Speranza for useful discussions. We also thank Dan Lee and Andr\'e Neves for discussing their work with us.
This work was supported by the STFC grant ST/L00044X/1. SF is supported by the ERC Advanced grant No.~290456.  AH is supported by an STFC studentship.

%
\bibliographystyle{apsrev4-1}
\bibliography{paperLocalErgbib}
%

\appendix

\section{Sub- and Supersolutions to IMC Flow}
\label{app:IMCF}

In this Appendix, we will generalise Lemma~3.1 of~\cite{LeeNev15} to the case where the asymptotic boundary of the asymptotically hyperbolic constant time slice~$\Sigma$ is not a space of constant curvature.  This allows us to apply the Weak Existence Theorem~3.1 of~\cite{HuiIlm01} to ensure that any weak IMC flow will asymptote to the boundary.

Let us write the induced metric on~$\Sigma$ in FG coordinates (in an arbitrary conformal frame):
\be
ds^2 = \frac{\ell^2}{z^2}\left[dz^2 + \left(\bar{g}_{\alpha\beta}(x) + O(z^2)\right) dx^\alpha dx^\beta\right],
\ee
where~$\alpha$,~$\beta$ label boundary spatial coordinates.

Then we claim that for any~$c > 0$, there exists a~$\zeta_0$ such that for all~$\zeta \leq \zeta_0$,
\be
\label{eq:subsup}
z^\pm(s) = \frac{1}{(1/\zeta \pm c) e^{s/2} \mp c}
\ee
are sub- and supersolutions to the IMC flow with initial condition~$z(0) = \zeta$.  To see this, first note that the mean curvature of a surface of constant~$z$ is
\bea
K(z) &= \frac{z^2}{2} \, \bar{g}^{ij} \pounds_{\hat{\partial}_z} \left(\frac{1}{z^2} \, \bar{g}_{ij}\right) \nonumber \\
				  &= \frac{2}{\ell} + O\left(z^2\right) \label{eq:Kpm},
\eea
where~$(\hat{\partial}_z)^a \equiv -(\partial_z)^a/\|\partial_z\|$ is the (outward-pointing) unit normal to these surfaces.  Thus the velocity of a flow starting on a~$z = \mathrm{const.}$ surface is initially~$K^{-1}(z) = \ell/2 + O(z^2)$.

Then note that the normal vectors to the claimed sub- and supersolutions above are
\bea
n_\pm^a &= \frac{dz^\pm}{ds} (\partial_z)^a \nonumber \\
		&= -\frac{z^\pm}{2}(1 \pm c z^\pm) (\partial_z)^a,
\eea
so that their velocity is
\be
\| n_\pm \| = \frac{\ell}{2} \left(1 \pm c z^\pm \right).
\ee
For small enough~$z$ (and thus~$\zeta$), comparing with~\eqref{eq:Kpm} we see that these will always be larger and smaller than~$K^{-1}(z^\pm)$, so these are indeed sub- and supersolutions.  Moreover, it then follows from~\eqref{eq:subsup} (or alternatively, from a direct integration of~\eqref{eq:Kpm}) that asymptotically, any solution to the weak IMC flow must behave as
\be
\label{eq:FGIMCF}
z = f(x) e^{-s/2} + \cdots,
\ee
where $\cdots$ represent terms that are subleading in~$s$.  Here~$f(x)$ satisfies~$(1/\zeta-c)^{-1} \geq f(x) \geq (1/\zeta+c)^{-1}$ for some~$\zeta > 0$,~$c > 0$, which in particular implies that~$f(x)$ is nowhere-vanishing.  To obtain~\eqref{eq:FGIMCF}, we assumed we could start the flow on a surface of constant~$z = \zeta$.  However, we may of course convert from one FG coordinate system to any other, with the leading-order transformation between the two FG radial coordinates~$(z,z')$ taking the form~$z' = \omega(x) z + O(z^3)$ for a nowhere-vanishing~$\omega(x)$.  Then~\eqref{eq:FGIMCF} remains unchanged, and is thus valid in any FG coordinate system.

\section{IMC Flow in Pure AdS}
\label{app:AdSIMCF}

Here we show explicitly how different IMC flows in pure AdS pick out different conformal frames.  We will consider IMC flows on a static time slice of pure AdS which start from a point; due to the maximal symmetry of AdS, the surfaces of all such flows will be spheres all the way to the conformal boundary.  We start with the metric of AdS in usual global coordinates,
\be
ds^2 = -(1+r^2/\ell^2) dt^2 + \frac{dr^2}{1+r^2/\ell^2} + r^2 \left(d\theta^2 + \sin^2\theta \, d\phi^2 \right),
\ee
and then perform a coordinate transformation to new variables~$(\tilde{r},\tilde{\theta})$:
\begin{subequations}
\bea
\frac{r^2}{\ell^2} + 1 &= \left(\sqrt{1+\frac{R^2}{\ell^2}}\sqrt{1+\frac{\tilde{r}^2}{\ell^2}} - \frac{R \tilde{r}}{\ell^2} \cos(\tilde{\theta})\right)^2, \\
\sin\theta &= \frac{\tilde{r}}{r} \sin\tilde{\theta},
\eea
\end{subequations}
where~$R$ is an arbitrary constant.  This transformation acts as an isometry on the slices of constant~$t$ and simply serves to shift their origin by a coordinate distance~$R$.  In these shifted coordinates, the metric is
\begin{multline}
ds^2 = -\left(\sqrt{1+\frac{R^2}{\ell^2}}\sqrt{1+\frac{\tilde{r}^2}{\ell^2}} - \frac{R \tilde{r}}{\ell^2} \cos(\tilde{\theta})\right)^2 dt^2 \\ + \frac{d\tilde{r}^2}{1+\tilde{r}^2/\ell^2} + \tilde{r}^2 \left(d\tilde{\theta}^2+\sin^2\tilde{\theta} \, d\phi^2\right).
\end{multline}

Because the slices of constant~$t$ are left unchanged, by spherical symmetry the spheres~$\tilde{r} = \mathrm{const.}$ will define an IMC flow that starts at~$\tilde{r} = 0$ and asymptotes to the conformal boundary~$\tilde{r} \to \infty$.  In terms of the original global coordinates, such a flow starts at~$(r,\theta) = (R,0)$, and thus each possible choice of~$R$ defines a different IMC flow.  Now,~$\widetilde{\Omega} = \ell/\tilde{r}$ is a conformal factor adapted to such a flow; then the boundary metric in the corresponding flow frame is
\begin{multline}
ds^2 = -\left(\sqrt{1+\frac{R^2}{\ell^2}} - \frac{R}{\ell} \cos (\tilde{\theta} )\right)^2 dt^2 \\ + \ell^2 \left(d\tilde{\theta}^2+\sin^2 \tilde{\theta} \, d\phi^2\right).
\end{multline}
The~$R$-dependence makes clear that different flows pick out different adapted conformal frames.

\end{document}